\documentclass[aps,prd,twocolumn,showpacs,preprintnumbers,nofootinbib,amsmath,amssymb]{revtex4}

\usepackage{graphicx}
\usepackage[caption=false]{subfig}
\usepackage{bm}
\usepackage{amsthm,url}
\usepackage{color,ulem}
\usepackage{cancel}
\usepackage{slashed}
\usepackage{lipsum}
\usepackage{float}
\usepackage{enumitem}

\def\be{\begin{equation}}
\def\ee{\end{equation}}
\def\bea{\begin{eqnarray}}
\def\eea{\end{eqnarray}}
\def\bn{\begin{enumerate}}
	\def\en{\end{enumerate}}
\def\bsube{\begin{subequations}}
	\def\esube{\end{subequations}}
\def\ll{\left}
\def\rr{\right}

\def\mc{\mathcal}
\def\mb{\mathbf}
\def\nn{\nonumber}
\def\ll{\left}
\def\rr{\right}

\def\h{\mb{h}}
\def\F{\mb{F}}
\def\x{\mb{x}}
\def\s{\mb{s}}

\def\z{\mb{z}}

\def\h{\mb{h}}
\def\Lc{\mc{L}}
\def\f{\text{F}}
\def\F{\mb{F}}
\def\Ro{\boldsymbol{\rho}}

\def\A{\text{A}}
\def\Bc{\mc{B}}

\def\Pr{\textbf{P}}

\usepackage{color}

\makeatletter

\newcommand{\Rmnum}[1]{\expandafter\@slowromancap\romannumeral #1@}
\makeatother

\begin{document}
	
	\preprint{------}
	
	\title{Performance comparison of multi-detector detection statistics in targeted compact  binary coalescence GW search} 
	\author{K Haris\footnote{Electronic address: haris@iisertvm.ac.in} and Archana Pai\footnote{Electronic address: archana@iisertvm.ac.in}}
	\affiliation{Indian Institute of Science Education and Research Thiruvananthapuram, CET Campus, Trivandrum 695016}
	\date{\today}
	\begin{abstract}
		Global network of advanced Interferometric gravitational wave (GW) detectors are expected
		to be on-line soon. Coherent observation of  GW from a distant compact  binary coalescence  (CBC)  with a network of interferometers located in different continents give crucial
		information about the source such as source location
		and polarization information. In this paper we compare different multi-detector network detection statistics for CBC search. In maximum likelihood ratio (MLR) based detection approaches, the likelihood  ratio is optimized to obtain the best model parameters and the best  likelihood ratio value is used as statistic to make decision on the presence of signal. However, an alternative Bayesian approach involves marginalization of the likelihood ratio over the parameters to obtain the average likelihood ratio. We obtain an analytical expression for the Bayesian statistic  using the two effective synthetic data streams for  targeted search of non-spinning compact binary systems with an uninformative prior on the parameters. Simulations are carried out for testing the validity of the approximation and comparing the detection  performance with the maximum likelihood ratio based statistics. We observe that the MLR {\it hybrid} statistic gives comparable or better performance with respect to the Bayesian statistic. 
	\end{abstract}
	
	
	\maketitle
	
	\section{Introduction}
	A new exiting era of gravitational wave (GW) astronomy is opened with the first direct detection of GW signal from a binary black hole merger event \cite{PhysRevLett.116.061102} by US based Advanced laser interferometric detectors LIGO - Hanford and LIGO - Livingston \cite{0264-9381-32-7-074001,Harry:2010zz}. The Advanced Virgo detector will join the network soon \cite{TheVirgo:2014hva,Avirgo}. With up coming detectors like Japanese cryogenic detector KAGRA \cite{PhysRevD.88.043007,0264-9381-29-12-124007} and US-Indian detector LIGO-India \cite{LIGOIndia}, the global network of broad band advanced GW detectors will be able explore the universe in GW window. 
	
	Compact binary coalescences (CBC) with neutron stars and black holes are prime sources of gravitational waves for the ground based advanced detector network. Based on LIGO's first detection, we expect to see 30 or more binary black hole mergers in the 9-month observing run in 2017-2018 \cite{PhysRevLett.116.061102}. By 2019, with design sensitivity, advanced LIGO detectors could additionally observe 40 neutron star binary events and 10 neutron star - black hole (NS-BH) events per year \cite{0264-9381-27-17-173001} along with a hundred or more binary black hole mergers. These numbers would further improve with the new detectors in global interferometric detectors.
	
		The classical detection procedure of GW involves defining a detection statistic, which is a function of data and is compared it with a threshold. If all the signal parameters are known, then by {\it Neyman-Pearson} lemma, the likelihood ratio $\textbf{Exp}[\Lambda(x)]$ - the ratio between  probabilities of hypotheses, $\mc{H}_S:$ the data $\x$ contains signal and  $\mc{H}_N:$ the data $\x$ contains  purely noise -  is the most powerful detection statistic \cite{nla.cat-vn4318735}.
		
		\be
		\textbf{Exp}[\Lambda(x)] = \frac{\Pr(x|\mc{H}_S)}{\Pr(x|\mc{H}_N)}~. \label{LR0}
		\ee
		%
		%
		
		However in GW detection problem, the signal parameters are unknown. i.e.  $\mc{H}_S$  is composite hypothesis rather than simple hypothesis. There are two distinct statistics used to  address  composite hypothesis testing problem \cite{nla.cat-vn4318735}.  
		
		\begin{itemize}
			\item {\it Maximum Log Likelihood Ratio(MLR) statistic, } $\Lc(x)$ is the maximum of the log of likelihood ratio in the multi-dimensional signal parameter space. If $ \boldsymbol {\hat \Theta}$ is the point at which the likelihood ratio in Eq.\eqref{LR0} is maximum, then
			
			\be
			\Lc(x) = \Lambda (x, \boldsymbol {\hat \Theta}) 
			\ee
			\item {\it Bayesian detection statistic or Bayes factor statistic,} $\mc{B}(x|I)$ is obtained by marginalizing the likelihood ratio over the parameter set $\boldsymbol{\Theta}$ with a prior distribution $\Pr(\boldsymbol{\Theta}|\mc{H}_s I)$. {\it i.e,}
			
			\bea
				\mathbf{Exp} [\mc{B}(x|I)] &=&  \int_{\boldsymbol{\Theta}} \frac{\Pr(x|\mc{H}_s(\boldsymbol {\Theta}) I)}{\Pr(x|\mc{H}_nI)}~ d \boldsymbol{\Theta} \nn \\
				&=& \int_{\boldsymbol{\Theta}} ~ e^{\Lambda(x;\boldsymbol{\Theta})}~ \Pr(\boldsymbol{\Theta}|\mc{H}_s I)~ d \boldsymbol{\Theta}~. \label{Bstat1}
			\eea
		\end{itemize}
		
		The multi-detector MLR approach for CBC signals has been developed in the literature by  various groups \cite{Pai:2000zt,Harry:2010fr,Haris:2014fxa}. In this paper we explore the Bayesian approach in the context of multi-detector CBC  search.  
	
	The paper is divided as follows. In Sec.\ref{detection}, we discuss  different detection statistics for multi-detector CBC search developed so far. We  show that the multi-detector  MLR statistic is a Bayesian statistic with an unphysical prior.  In Sec.\ref{Bphysical}, we derive an approximate analytic expression for  multi-detector  Bayesian detection statistic. In Sec.\ref{SecB0pi}, we derive  the  Bayesian detection statistic  tuned for face-on/off binaries and construct a {\it hybrid} statistic. In Sec.\ref{simulations}, we assess the performance of the statistics by comparing the Receiver Operator Characteristic (ROC) curves.  Finally in Sec.\ref{conclussion},  we summarize the conclusion.

	\section{MLR detection approaches for CBC search}\label{detection}
	
		The GW signal from a non spinning compact binary source such as, double neutron stars or neutron star - black hole binaries with negligible spin, are characterized by a set of 9 parameters, $\{m_1,m_2,\A,t_a,\phi_a,\epsilon,\Psi,\theta,\phi\}$, where $(m_1, m_2)$ are the component masses and $\A$ is the constant overall amplitude. $\phi_a$ is the phase of the waveform at the time of arrival $t_a$, $(\epsilon,\Psi)$ are the  inclination angle and  polarization angle respectively. $(\theta,\phi)$ characterizes the location of the source in the celestial globe in geocentric coordinates. The time domain GW signal at any $m^{th}$ detector can be written as\cite{Haris:2014fxa},
		{\small \be
			s_m(t) = \f_{+m} h_+(t) +  \f_{\times m}\h_\times (t)~,
			\ee}
		
		where the antenna patterns $\f_{+m}$ and $\f_{\times m}$ are functions of source location $(\theta,\phi)$ and the detectors Euler angles in  geocentric coordinates. A detailed description of coordinates are given in \cite{apaiprd06}. The 3.5 PN restricted non-spinning GW polarizations $\h_+ (t)$ and $\h_\times (t)$ are functions of $\{m_1,m_2,\A,t_a,\phi_a,\epsilon,t_a\}$. The parameters $\A, \epsilon, \Psi, \theta, \phi, t_a $ and  $\phi_a$ appear in  the signal as either a scale or a time/frequency shift. Hence they are termed as {\it extrinsic} parameters. The phase evolution of waveform is characterized by the  masses $\{m_1,m_2,\}$, which are termed as {\it intrinsic parameters}.
		
		For a global network of $I$ interferometric detectors with uncorrelated noise, the optimum network matched filter SNR square can be expressed as sum of squares of SNRs of individual detectors. {\it i.e}\footnote{The scalar product of $\mb{a}$ and $\mb {b}$ is defined as  $$\langle\mb{a}|\mb{b}\rangle = 4 \Re \ll[ \int_0^\infty \tilde{\tilde a}(f)~ \tilde b^*(f)~ df \rr],$$ where $\tilde{\tilde{a}}(f) = \tilde a(f)/S_n(f)$ is the double-whitened version of frequency series $\tilde  a(f)$. The $S_n(f)$ is the one sided noise power spectral density(PSD) of a detector. In discrete domain, $$\langle\mb{a}|\mb{b}\rangle = 4 \Re \ll[ \sum_{j=1}^N \tilde{\tilde a}_j~ \tilde b^*_j \rr],$$ where $j$ is the frequency index.},
		
		\be
		\Ro_s^2 = \sum_{m=1}^I \langle \s_m | \s_m\rangle. \label{SNR}
		\ee
			The corresponding log likelihood ratio for Gaussian noise is given by,
			{\small \be
				\Lambda(X) = \sum_{m=1}^I \langle \x_m| \s_m \rangle -\frac{1}{2} \langle  \s_m| \s_m \rangle.
				\ee}

	Maximization or marginalization of likelihood ratio over extrinsic parameters, $\{A, \phi_a, \cos \epsilon, \Psi\}$  can be done in a straight forward fashion compared to intrinsic parameters. As intrinsic parameters alter the shape of the waveform, maximization/marginalization is a numerical problem. The statistics, which are obtained by either maximizing or marginalizing likelihood ratio over extrinsic parameters, is then used to search over the remaining intrinsic parameters. 
	
	In the following sections we derive and compare MLR and Bayes factor statistics for targeted non spinning inspiral search with multi-detector network. 
	\subsection{Review of coherent multi-detector MLR statistic}
	
	Coherent multi-detector network MLR analysis for CBC signal  was formulated in literature by various groups\cite{Pai:2000zt,Harry:2010fr,Haris:2014fxa}. In this section we review the same.
	
	 The log of network likelihood ratio  for $I$ interferometric detector in the network with uncorrelated Gaussian noises can be written in terms of a pair of synthetic streams as \cite{Haris:2014fxa},
	{\small \be
		\Lambda(X) ~=~ \ll[\boldsymbol{\rho_L} \langle  \z_L|   \h_0 e^{i  \Phi_L} \rangle - \frac{ \boldsymbol{\rho_L}^2}{2} \rr] +\ll[\boldsymbol{\rho_R} \langle  \z_R|  \h_0 e^{i  \Phi_R} \rangle  - \frac{\boldsymbol{\rho_R}^2}{2} \rr] ~, \label{LR}
		\ee}
	Here the over-whitened synthetic streams  $\tilde{\tilde \z}_{L,R}$ are construed as a linear combination of  over-whitened  data, $\tilde {\tilde \x}_{m}$ from individual detectors  as below,
	
	\be \tilde {\tilde z}_L(f) =\sum_{m=1}^I \frac{{\text{F}}_{+m}}{\| \F'_+\|} \tilde {\tilde x}_{m}(f)~,
	~~~ \tilde {\tilde z}_{R}(f) =\sum_{m=1}^I \frac{{\text{F}}_{{\times}m}}{\| \F'_{\times}\|} \tilde {\tilde x}_{m}(f)~. \label{eq:z1z2}
	\ee
	
	$\F = \F_{+} + i \F_{\times}$ is the  complex network antenna pattern vector in dominant polarization frame and $\F' = \F'_{+} + i \F'_{\times} \equiv \{g_m \textbf{F}_m\}$ is the noise weighted versions of the same, with the noise weight $g_m  = \langle \h_0| \h_0 \rangle_m$. The dominant polarization  frame is the wave frame in which the plus and cross noise weighted antenna pattern vectors are orthogonal to each other in the network space \cite{PhysRevD.72.122002}. The dominant polarization frame allows the $\Lambda(X)$ to be written as a sum of  log likelihood ratios of a pair of synthetic  detectors obtained {\it via} synthetic streams.

	The relation between the new set of derived extrinsic parameters, $\{ \Ro_L,\Ro_R,\Phi_L,\Phi_R \}$ and physical extrinsic parameters, $\{\text{A}_0, \phi_a, \cos \epsilon, \Psi \}$ is given in Appendix-\ref{appendix-extrinsic}. $\Ro_{L,R}$ and $\Phi_{L,R}$ act as the SNRs and overall phases of  two effective synthetic detectors respectively.

	The  maximum log likelihood ratio, $\Lc(X)$ maximized over these derived parameters  is then  a sum of quadratures of the two synthetic streams as below \cite{Harry:2010fr,Haris:2014fxa},
	
	{\small \bea
		\Lc (X) = \underbrace{\langle  \z_L |  \h_0 \rangle^2 + \langle  \z_L |  \h_{\pi/2} \rangle^2}_{\hat{\Ro}^2_L} 
		+\underbrace{\langle  \z_R |  \h_0 \rangle^2 + \langle  \z_R |  \h_{\pi/2} \rangle^2}_{\hat{\Ro}_R^2} ~, \label{MLR}
		\eea}
	where $\tilde{\h}_0$ and $\tilde{\h}_{\pi/2} = -i \tilde{\h}_0 $ are the two GW phases..
	
	$\hat \Ro_{L,R}$ are the maximum likelihood estimates of $\Ro_{L,R}$, the optimum SNRs of the synthetic streams. The estimates of  $\Phi_{L,R}$  are given by,
	
	{\small \be
		\hat {\Phi}_{L,R} = arg \ll[\sum_{j=1}^N \tilde {\tilde z}_{L,R j}~ \tilde h^*_{0j} \rr]~.
		\label{MLREstimates}
		\ee}

	\subsection{MLR as viewed in Bayesian framework}\label{MLRasB}
	 In this subsection we show that the MLR statistic, $\Lc$ can be understood as a $\Bc$ statistic with an unphysical prior, $\Pi_{unph}$ over the  extrinsic parameters. The likelihood ratio is expressed in terms of  $\{ \Ro_L,\Ro_R,\Phi_L,\Phi_R \}$ (see Eq.\eqref{LR}).
	
	If we choose,
	
	{\small \be
		\Pr(\Ro_{L,R},\Phi_{L,R} |\Pi_{unph} ) = C \Ro_L \Ro_R ; \begin{cases}
			\Ro_{L,R} &\in [0,\infty),\\
			\Phi_{L,R} &\in [0,2\pi).
		\end{cases} \label{Pi_c1}
		\ee}
	as the prior distribution for these parameters, then a closed form expression for the  Bayesian statistic can be obtained in a straight forward way, with $C$ as a normalization constant. Since $\Ro_{L,R}$ are the synthetic SNRs and $\Phi_{L,R}$ are the effective phases, we allow them to take values in the entire range.
	
	Substituting Eq.\eqref{LR}, Eq.\eqref{MLR} and Eq.\eqref{MLREstimates} in Eq.\eqref{Bstat1} we get,
	
	{\small \bea
		\mathbf{Exp} [\Bc(X|\Pi_{unph})] &=& C \int_0^\infty d \Ro_L  \int_0^{2 \pi} d \Phi_L~ \Ro_L~ e^{\Ro_L \hat{\Ro}_L \cos (\Phi_L - \hat{\Phi}_L)} \nn \\
		 \int_0^\infty  d \Ro_R&&\int_0^{2 \pi} d \Phi_R~  \Ro_R~ e^{\Ro_R \hat{\Ro}_R \cos (\Phi_R - \hat{\Phi}_R)}~. \label{B_c}
		\eea}
	By rearranging terms, the integral Eq.\eqref{B_c} can be converted into a product of four Gaussian integrals and thus the statistic finally becomes,
	\bea
	\mathbf{Exp} [\Bc(X|\Pi_{unph})] ~= ~C'~ e^{\frac{\hat{\Ro}_L^2 + \hat{\Ro}_R^2}{2}} ~=~C'~ e^{\Lc (X) /2},
	\eea
	with $C' = 4 \pi C~$, or
	
	\be
	\Bc(X|\Pi_{unph})  = \frac{\Lc(X)}{2} + C'. \label{Bstat0}
	\ee
	  Eq.\eqref{Bstat0} clearly indicates that the maximum log likelihood ratio $\Lc(X)$ is proportional to the Bayesian statistic, $\Bc(X)$ with a  prior $\Pi_{unph}$. In other words, the maximum likelihood ratio is a Bayesian statistic in this prior $\Pi_{unph}$. 
	
	To understand the physical meaning of the prior $\Pi_{unph}$, we  obtain corresponding probability distribution of physical  parameters $\{A, \phi_a, \cos \epsilon, \Psi\}$. From Eq.\eqref{Pi_c1}, the probabilities  of  physical  parameters is given by,
	{ \small \bea
		\Pr(A, \phi_a, \cos \epsilon, \Psi | \Pi_{unph}) &=& C~ \Ro_L \Ro_R~ |J| \nn \\
		&=& \frac{\A^3~ \|\F_+'\|^2 \|\F_\times'\|^2}{4}~ (1-\cos^2 \epsilon)^3, \nn \\  \label{Pi_c2}
		\eea}   
	where $|J|$ is the determinant of Jacobian of transformation from parameter set $\{\Ro_L,\Phi_L, \Ro_R,\Phi_R\}$ to $\{A, \phi_a, \cos \epsilon, \Psi\}$. i.e,
	\bea
	|J| &=&\ll| \det\ll(\frac{\partial \{\rho_L,\Phi_L,\rho_R,\Phi_R\} }{\partial \{\text{A}_0, \phi_a, \cos \epsilon, \Psi\} } \rr)\rr| \nn \\  &=& \frac{\A_0^3 ~\|\F_+'\|^2~ \|\F_\times'\|^2~ (1-\cos^2 \epsilon)^3}{4~ \Ro_L(A, \phi_a, \cos \epsilon, \Psi) ~\Ro_R(A, \phi_a, \cos \epsilon, \Psi)}~.
	\eea
	See Appendix.\ref{appendix-extrinsic} for details. Close look at  Eq.\eqref{Pi_c2} shows that the assumed prior distribution in Eq.\eqref{Pi_c1}  is more biased towards the Edge-on case compared to face-on case. In reality, we expect that more observations from face-on/off systems due to high SNR compared to the edge-on. A similar observation was made in \cite{0264-9381-26-20-204013} for the case of continuous wave sources, where the connection between the MLR statistic and $\Bc$ statistic was first obtained in the literature. 
	
	\section{ Bayesian Statistic for a physical prior, $\Pi_{ph}$} \label{Bphysical}
	
	
	In this section, we derive a $\Bc$ statistic for a physical prior.  Since we don not have any prior information on any of the parameters, we use  flat (uninformative)   prior for the physical parameters. Using reasonable approximations we solve the integral in Eq.\eqref{Bstat1}, to obtain  $\Bc$ statistic. Further we test the validity of the approximation used. In the remaining part of the paper the notation $\Bc$ represents the Bayesian statistic with physical prior, $\Pi_{ph}$ unless specified otherwise. 
	
	\subsection{Physical Prior, $\Pi_{ph}$}\label{sec-Phyprior}
	The inclination angle, $\epsilon$ and polarization angle, $\Psi$  together form a spherical polar coordinate pair in {\it polarization sphere},  in which $\epsilon$ will act as the polar angle and $\Psi$ act as corresponding azimuthal angle. Sampling points uniformly from the spherical surface is the most natural uninformative prior for $(\epsilon,\Psi)$. We note that $\F_{+,\times}(\Psi) = - \F_{+,\times}(\Psi + \frac{\pi}{2})$. i.e, $2 \Psi$ has  $\pi$ symmetry in the GW signal. Therefore, 
	\begin{alignat}{3}
		\Pr(\cos \epsilon|\Pi_{ph}) &=&~ \frac{1}{2 }~;~~~~~~~~~ \cos \epsilon &\in [-1, 1]~,& \nn \\
		\Pr(\Psi|\Pi_{ph}) &=&~ \frac{2}{ \pi}~;~~~~~~~~~~~~~ \Psi &\in [-\frac{\pi}{4}~,& \frac{\pi}{4})~. \label{prior_ph1}
	\end{alignat}
	
	The probability distributions of the amplitude $A$ and the initial phase $\phi_a$ are chosen to be  uniform for simplicity. {\it i.e,}
	\begin{alignat}{3}
		\Pr(\phi_a|\Pi_{ph}) &=&~ \frac{1}{2 \pi}~; ~~~~~~~~~~~~   \phi_a &\in [0, 2 \pi]~,& \nn \\
		\Pr(\text{A}_0|\Pi_{ph}) &=&~ \frac{1}{\text{A}_0^{max}}~;~~~~~~~ \text{A}_0 &\in [0, \text{A}_0^{max}]~,& \label{prior_ph2}
	\end{alignat}
	where $\text{A}_0^{max}$ is the upper limit for the amplitude.
	
	Thus the combined prior distribution is,
	{\small \be
	\Pr(\text{A}_0, \phi_a, \cos \epsilon, \Psi|\Pi_{ph}) = \frac{1}{2 \pi^2 \text{A}_0^{max}} \equiv C''~.\label{prior_ph3}
	\ee}

Using $|J|$ in Eq.\eqref{jacobian}, we obtain the corresponding distribution of the new  extrinsic parameters,  $\Pr(\Ro_L,\Ro_R,\Phi_L,\Phi_R|\Pi_{ph})$  as,
	
	{\small \be
		\Pr(\Ro_L,\Ro_R,\Phi_L,\Phi_R|\Pi_{ph}) = \frac{2 C'' \Ro_L ~\Ro_R~\ll|\frac{\Ro_L^2 e^{-2i\Phi_L}}{\|\F'_+\|^2} + \frac{\Ro_R^2 e^{-2i\Phi_R}}{\|\F'_\times\|^2}\rr|^{-\frac{3}{2}}}{\|\F_+'\|^2 ~\|\F_\times'\|^2 }~. \label{prior_ph4}
		\ee}
	
	If $\ll|\frac{\Ro_L^2 e^{-2i\Phi_L}}{\|\F'_+\|^2} + \frac{\Ro_R^2 e^{-2i\Phi_R}}{\|\F'_\times\|^2}\rr| = 0$   , the probability distribution, $\Pr(\Ro_L,\Ro_R,\Phi_L,\Phi_R|\Pi_{ph})$ in   Eq.\eqref{prior_ph4} diverges. For that case, the determinant of  Jacobian  $|J|$ vanishes. {\it i.e}, The transformation between $\{\text{A}_0, \phi_a, \cos \epsilon, \Psi\}$ and $\{\Ro_L,\Ro_R,\Phi_L,\Phi_R\}$ is invalid. This  happens for face-on/off case, where $\frac{\rho_L}{\|\F_+'\|} = \frac{\rho_R}{\|\F_\times'\|}$, and  $\hat{\Phi}_L = \hat{\Phi}_R \pm \frac{\pi}{2}$. In this cases, the GW becomes circularly polarized, where $\Psi$ and $\phi_a$ become degenerate. We exclude this case from the below derivation of  $\Bc$ statistic. We treat face-on/off as a special case and obtain the $\Bc$ statistic for face-on/off in Sec.\ref{SecB0pi}.
	
	\subsection{Bayesian Statistic, $\Bc(X|\Pi_{ph})$}
	
	In this subsection, we derive an analytic approximation for the  $\Bc$ statistic. We substitute  Eq.\eqref{prior_ph3} in  Eq.\eqref{Bstat1} and assume $\text{A}_0^{max} >> 1$, this gives the Bayesian statistic as, 
	{\small \bea
		\mathbf{Exp} [\mc{B}(X|\Pi_{ph}I)] &=& C'' \int_0^\infty d\text{A}_0 \int_{-1}^1 d \cos \epsilon \nn \\ && \int_{-\pi/4}^{\pi/4} d \Psi \int_0^{2 \pi} d \phi_a~ e^{\Lambda(X)}~.  \label{Bstat_1}
		\eea}
	
	Using Eq.\eqref{prior_ph4}, we can rewrite  Eq.\eqref{Bstat_1} in terms of $\{\Ro_L,\Ro_R,\Phi_L,\Phi_R\}$  as,
	{\small  \begin{widetext}
			\bea
			\mathbf{Exp} [\mc{B}(X|\Pi_{ph}I)] &=& C'' \int_0^{\infty} d\rho_L \int_0^{\infty} d \rho_R \int_0^{2 \pi} d \Phi_L \int_0^{2 \pi} d \Phi_R~  \frac{e^{\Lambda(X)}}{|J|}~ \nn \\
			&=& \frac{C''}{\|\F_+'\|  \|\F_\times'\|} \int_0^{\infty}  d\rho_L \int_0^{\infty} d \rho_R \int_0^{2 \pi} d \Phi_L \int_0^{2 \pi} d \Phi_R~ \rho_L \rho_R~ \ll| \frac{\rho_L^2}{\|\F_+'\|^2} e^{2 i \Phi_L} + \frac{\rho_R^2}{\|\F_\times'\|^2} e^{2 i \Phi_R} \rr|^{-\frac{3}{2}} \nn \\
			&& \text{Exp}\ll[\rho_L\hat{\rho}_L \cos (\Phi_L - \hat{\Phi}_L) - \frac{1}{2} \rho_L^2\rr]~ \text{Exp}\ll[\rho_R \hat{\rho}_R \cos (\Phi_R - \hat{\Phi}_R)  - \frac{1}{2} \rho_R^2\rr] ~. \label{Bstat_2}
			\eea
		\end{widetext}}

		We note that, though the numerator is separated in $\{\rho_L,\Phi_L\}$ and $\{\rho_R,\Phi_R\}$, the denominator is inseparable. In Eq.\eqref{Bstat_2} the numerator contains a product  of  exponential functions in $\{\Ro_L,\Ro_R,\Phi_L,\Phi_R\}$. Compared to this exponential term all the remaining terms, namely the denominator and $\Ro_L \Ro_R$ in the numerator  vary slowly in the parameter range. Further,  the product of exponential terms together have a single maximum at the maximum likelihood point, $\{\hat{\Ro}_L,\hat{\Ro}_R,\hat{\Phi}_L,\hat{\Phi}_R\}$. Assuming the denominator is stationary (slowly varying) around the maximum likelihood point and using Gaussian integral approximation, we can approximate   the integral in Eq.\eqref{Bstat_2} as,
		
		{\small \be
			\mathbf{Exp} [\mc{B}(X|\Pi_{ph})]
			=  \frac{2 \pi^2~ C''  ~e^{\frac{\hat{\rho}^2_L + \hat{\rho}^2_R}{2}} ~ \ll|  \frac{\hat{\rho}_L^2}{\|\F_+'\|^2 }   e^{2 i \hat{\Phi}_L} + \frac{ \hat{\rho}_R^2 }{\|\F_\times'\|^2 }  e^{2 i \hat{\Phi}_R} \rr|^{-\frac{3}{2}}}{\|\F_+'\| ~\|\F_\times'\| }~. \label{Integral6}
			\ee}
		
		Or, 
		{\small \bea
			\mc{B}(X) &=& \frac{\Lc(X)}{2} - \frac{3}{2} \ln  \ll| \frac{\hat{\rho}_L^2~e^{2 i \hat{\Phi}_L} }{\|\F_+'\|^2}  + \frac{\hat{\rho}_R^2~e^{2 i \hat{\Phi}_R}}{\|\F_\times'\|^2}  \rr| + const. \nn \\ \label{LogB1}
			\eea}
		\begin{figure*}
							\centering
							\subfloat{\includegraphics[width=1.\textwidth]{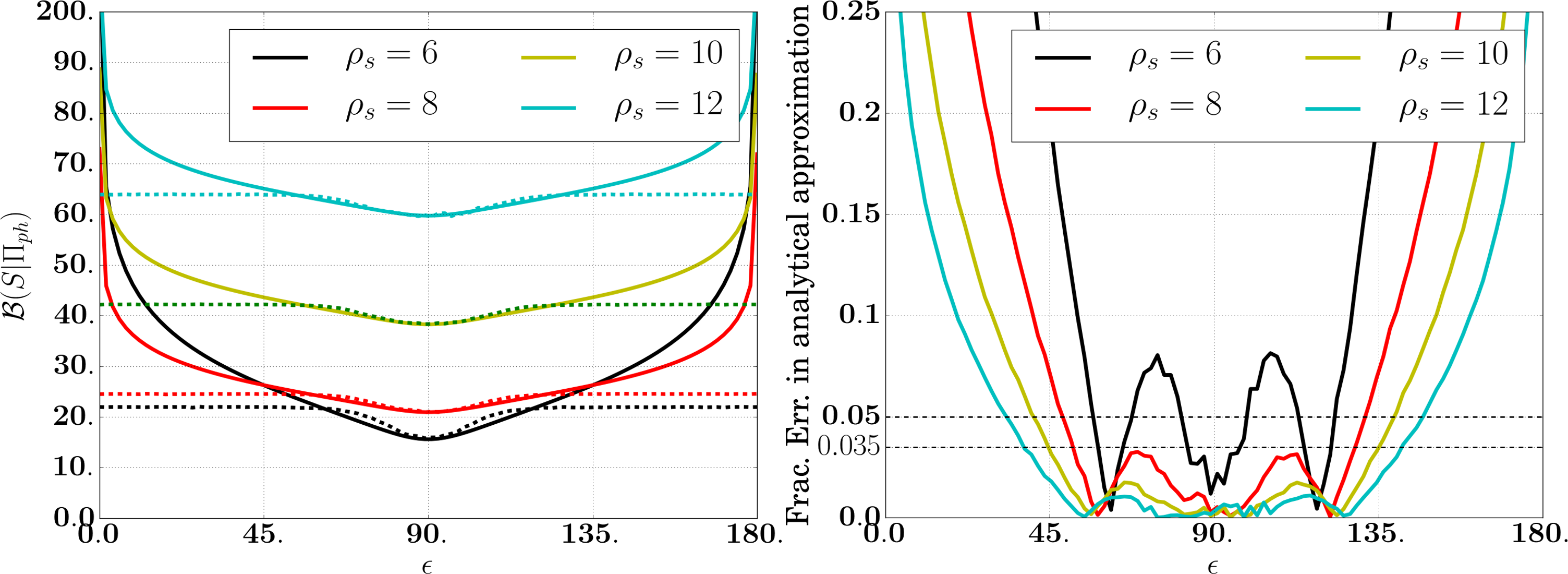}}
							\caption{\label{Stats} (a) Plots of $\Bc(X|\Pi_{ph})$ along inclination angle $\epsilon$ computed numerically (dashed curves) and analytically (solid cures) while the  data is purely signal for different optimum SNR $\Ro_s$. (b) Variation of fractional error in analytic approximation along $\epsilon$. The signal is from $(2. - 10. M_\odot)$ NS-BH system optimally located at $(\theta=140^\circ,\phi=100^\circ)$ with an arbitrary polarization angle $\psi = 45^\circ$. We assume "zero-detuning, high power" Advanced LIGO PSD\cite{aLIGOSensitivity} for all detectors.}
		\end{figure*}

	The detailed derivation of this integral  is given in Appendix.\ref{app-2}.
		
		To summarize, the approximation depends on two  conditions. 
		\begin{enumerate}[label=(\alph*)]
			\item The denominator of the integrand is not equal to zero around the maximum likelihood point. i.e, the GW signal is not from a face-on/off binary.  As mentioned earlier we treat this case in Sec.\ref{SecB0pi}.
			
			\item  Both synthetic stream matched filter SNRs $\hat{\Ro}_L$ and $\hat{\Ro}_R$, are reasonably high, else the corresponding Gaussian integral assumption breaks. 
		\end{enumerate}

		\subsection{$\Bc$ statistic in  $\{ \text{A}_L, \text{A}_R \}$ coordinates}
		
		In this subsection, we re-express the $\Bc$ statistic in Eq.\eqref{LogB1} in terms of a pair of amplitude coordinates namely, $\text{A}_{L,R} \equiv \text{A} ~(1 \pm \cos \epsilon)/2$ instead of $\{\Ro_L,\Ro_R,\Phi_L,\Phi_R\}$.

		From Eq.\eqref{coordinates}, we can relate $\{\Ro_L,\Ro_R,\Phi_L,\Phi_R \}$ to $\{ \text{A}_L, \text{A}_R \}$ as,
		
		{\small \bea
			\|\F_\times'\|^2 \rho^2_L e^{2 i \Phi_L} + \|\F_+'\|^2 \rho^2_R e^{2 i \Phi_R} 
			\equiv \|\F_+'\|^2 \|\F_\times'\|^2 \text{A}_L \text{A}_R, \nn \\ \label{LogB2}
			\eea}
		
		 Then by substituting Eq.\eqref{LogB2} in Eq.\eqref{LogB1}$,  \mc{B}(X|\Pi_{ph})$ in terms of $\text{A}_{L,R}$ becomes,
		
		\bea
		\mc{B}(X|\Pi_{ph}) &=&  \frac{\hat{\rho}^2_L+ \hat{\rho}^2_R }{2} - \frac{3}{2} \ln \ll[ \|\F_+'\|^2  \|\F_\times'\|^2 \hat{\A}_L \hat{\A}_R \rr]+ const. \nn \\
		&=& \frac{\Lc(X)}{2} - \frac{3}{2} \ln \ll[ \|\F_+'\|^2  \|\F_\times'\|^2 \hat{\A}_L \hat{\A}_R \rr] + const. \nn \\
		\eea

		This is an alternative representation of the statistic in terms of $\text{A}_{L,R}$. In \cite{0264-9381-26-20-204013}, the authors obtained a closed form expression for $\mc{B}$ statistic with prior $\Pi_{ph}$ in the continuous waves search context   by Taylor expanding the likelihood ratio about it's maximum value.  	The above equation  is similar  to Eq.(5.36) of \cite{0264-9381-26-20-204013}, which is obtained for continues GW source context.
		
		\subsection{Validity of the approximation}\label{validity}
		
	As described in the previous subsection, the validity of the analytical approximation crucially depends on two conditions. The first one is that the SNR of the signal should be high and the second one is that the source should not be face-on or edge-on. 
	
	In Fig.(\ref{Stats}), we have plotted the variation of $\Bc(X|\Pi_{ph})$ along the inclination angle, $\epsilon$ of the source for various network SNRs in the absence of noise to get an idea of the validity of the analytical expression. The numerical $\Bc$ statistic is obtained by Monte-Carlo (MC) simulation. We used $10^6$ MC points for the simulation.
	
	 For all SNR values, the fractional error diverges as the inclination angle, $\epsilon$  approaches, $0^\circ$ or $180^\circ$ (face-on/off). This is expected as determinant of $J$ is zero at these points. 
	 
	 From Fig.(\ref{Stats}.b) it is evident  that the fractional error in the analytic approximation reduces as the matched filter SNR of network increases. For example, at $\epsilon = 45^\circ$, the percentage fractional errors corresponding to $\Ro_s = 12, 10, 8, 6$ are $1.8\%, 3.1\%, 7\%$ and $19\%$  respectively. For a given fractional error, as the SNR increases the validity region of the  analytic expression increases. For a  percentage fractional error of $5 \%$, validity regions for $\Ro_s = 12, 10, 8$ are $\epsilon \in (48^\circ, 131^\circ)$, $\epsilon \in (40^\circ, 139^\circ)$ and $\epsilon \in (33^\circ, 147^\circ)$ respectively.
	 
		\section{Detection statistics for face-on/off binaries}\label{SecB0pi}
		
		We recall from previous section that the $\Bc$ statistic obtained with prior $\Pi_{ph}$ is invalid for face-on/off case. 
		This is because, in face-on/off case, the signal becomes purely circularly polarized and thus the polarization angle and initial phase are indistinguishable from each other\cite{Haris:2015iin}. 
		Explicitly in DP frame, at $\epsilon = 0,\pi$ the extrinsic parameters in Eq.\eqref{coordinates} can be reduced to,
		\bea
		\Ro_L &=& \text{A}_0 \|\F'_+\|~,~~~~ \Ro_R = \text{A}_0 \|\F'_\times\| \nn \\
		\Phi_L &=& \chi + \phi_a~,~~~~~~~~ \Phi_R = \chi + \phi_a \mp \frac{\pi}{2}, \label{coordinates_face}
		\eea
		where $\chi \equiv \Psi + \delta/4$. The angle $\delta$ is a function of source location and the multi-detector network configuration on Earth. For the detailed description of $\delta$, please refer \cite{Haris:2014fxa}.  The amplitudes $\Ro_{L,R}$ are constant times the signal amplitude $\text{A}_0$ and phase $\Phi_{R}$ is $\pi/2$ out of phase with $\Phi_L$. The polarization angle $\Psi$ and the initial phase $\phi_a$ are degenerate. Due to this degeneracy the log likelihood ratio for face-on/off binaries can be expressed in terms of two effective parameters $\{\Ro, \Phi_L\}$ as \cite{Haris:2015iin},
		
		\be
		\Lambda^{0,\pi}(X) =  \Ro~ \langle \z^{0,\pi}|\h_0 e^{i\Phi_L} \rangle - \frac{1}{2}\Ro^2~, \label{LLR-0pi} 
		\ee
		with $\Ro \equiv \A \|\F^{'}\|$ and 
		{\small \be
			\tilde {\tilde z}^0(f)  \equiv \sum_{m=1}^I \frac{\f_m}{\|\F^{'}\|} \tilde{\tilde x}_m(f)~,~~~ 
			\tilde {\tilde z}^\pi(f)  \equiv \sum_{m=1}^I \frac{\f^{*}_m}{\|\F^{'}\|} \tilde{\tilde x}_m(f)~ . \label{Z0Zpi}  
			\ee}
		\subsection{Maximum Likelihood  Statistic }
		The maximization of $\Lambda^{0,\pi}(X)$ over $\Ro$ and $\Phi_L$ is straight forward and the maximum likelihood ratio, $\Lc^{0,\pi}(X)$ is given by \cite{Williamson:2014wma,Haris:2015iin} as,
		
		\be
		\Lc^{0,\pi} (X)  = \langle \z^{0,\pi}|\h_0 \rangle^2 + \langle \z^{0,\pi}|\h_{\pi/2} \rangle^2, \label{MLR_0pi} 
		\ee
		with
		{\small \be
			\hat \rho = 4 \ll| \sum_{j=1}^N \tilde {\tilde z}^{0,\pi}_j~ \tilde h_{0j}^*  \rr|~, 
			~~\hat {\Phi}_L = arg \ll[\sum_{j=1}^N \tilde {\tilde z}^{0,\pi}_j~ \tilde h^*_{0j} \rr]~,
			\label{MLREstimates_0pi}
			\ee}
		as the maximum likelihood estimates of $\Ro$ and $\Phi_L$.

		In contrast with 2 stream generic MLR statistic, $\Lc(X)$, the MLR statistics tuned for face-on/face systems, $\Lc^{0,\pi}(X)$  are single stream statistics and  reduces the false alarm rate.  Further either $\Lc^0 (X)$ or $\Lc^{pi} (X)$ capture more than $98\%$ of network matched filter SNR for a wide range of binary inclination angles and polarization angles.  Because of above two properties, a new {\it hybrid}  statistic, $\Lc^{mx}(X)$ is proposed, which is the maximum of $\Lc^0 (X)$  , $\Lc^\pi (X)$ and gives better performance for a wide range of binary inclinations and polarizations \cite{Haris:2015iin}.
		
		\subsection{Bayesian statistic}
		In this subsection, we marginalize the likelihood ratio tuned for face-on/off binaries over $\Ro$ and $\Phi_L$ with the physical prior, $\Pi_{ph}$ discussed in Sec.\ref{sec-Phyprior}. 		For face-on/off binaries, the physical prior $\Pi_{ph}$ in Eq.\eqref{prior_ph2} reduces to, 
		
		\be
		\Pr(\Ro,\Phi_L|\Pi_{ph}) = \frac{1}{2 \pi \Ro^{max} },
		\ee
		with $\Ro^{max} \equiv \textbf{A}_0^{max} \|\F'\|$. Using Eq.\eqref{Bstat1}, Eq.\eqref{LLR-0pi} and Eq.\eqref{MLREstimates_0pi}, the Bayesian  statistic for face-on/off binaries can be written as,
		
		{\small \bea
			\mathbf{Exp} [\mc{B}^{0,\pi}(X)]  &=& \int_0^{2 \pi} d \Phi_L  \int_{ 0}^{\Ro^{max}} d \Ro \frac{e^{\Ro \hat{\Ro} \cos(\Phi_L-\hat{\Phi}_L) - \frac{1}{2} \Ro{2}}}{2 \pi \Ro^{max} } \nn \\
			&\approx& \frac{1}{\sqrt{2 \pi} \Ro^{max} }  \int_{0}^{2 \pi}  e^{\frac{\hat{\Ro}^{2} \cos^2(\Phi_L -\hat{\Phi}_L)}{2} }  ~d \Phi_L~, \label{Bstat_face1}
			\eea}
		provided $\hat{\Ro} \cos(\Phi_L -\hat{\Phi}_L)$ is not close to zero.  This condition is reasonable and  be satisfied for signals with high SNR, since the integrand will be significant only in a small window of $\Phi_L$ around $\hat{\Phi}_L$ . 
		
		We can approximate $\cos(\Phi_L - \hat{\Phi}_L)$  by $1-\frac{(\Phi_-\hat{\Phi}_L)^2}{2}$ and approximate the $\Phi_L$ integral by a Gaussian integral. Thus $\Bc^{0,\pi}$ can be approximated as,
		
		{\small \bea
			\mathbf{Exp} [\Bc^{0,\pi}(X)] &\approx&  \frac{\text{Exp}~\ll[\frac{\hat{\Ro}^{2} }{2}\rr]}{\hat{\Ro} \Ro^{max}  }~,  \label{Bstat_face2}
			\eea}
		
		and
		
		{\small \bea
			\mc{B}^{0,\pi}(X) &=& \frac{\hat{\Ro}^{2} }{2} - \ln \ll(\hat{\Ro} \Ro^{max}\rr)~, \nn \\
			&=& \frac{1}{2}~\Lc^{0,\pi}(X) - \ln \ll[\Lc^{0,\pi}(X)~\Ro^{max}\rr]~.  \label{Bstat_face3}
			\eea}
		
		This implies  $\mc{B}^{0,\pi}$ statistic can be approximated by MLR statistic $\Lc^{0,\pi}$ with a small logarithmic correction, $\ln \ll[\Lc^{0,\pi}(X)~\Ro^{max}\rr]$. In the same spirit of $\Lc^{mx} (X)$, we can define a {\it hybrid}  Bayesian statistic $\Bc^{mx}$ as,
		
		\be
		\Bc^{mx} (X|\Pi_{ph}) = \max\ll[\Bc^0(X) , \Bc^\pi (X)\rr]. 
		\ee
		
		\section{Simulations and Discussion} \label{simulations}
		In this section, we carry out numerical simulations for  three detector network LHV, with Ligo-Livingston (L), Ligo- Hanford (H)
		and Virgo (V) as the constituent detectors to test the validity of analytical $\Bc$ statistics and compare the performance of detection statistics by means of Receiver Operator Characteristic (ROC) curve, which is the plot between false alarm probability (FAP) and detection probability (DP). All the detectors are assumed to have additive Gaussian random noise with the noise PSD following ”zero-detuning, high power” Advanced LIGO noise curve\cite{aLIGOSensitivity}.  The simulations are performed for $(2 - 10 M_\odot)$ NS-BH non spinning binary signal with fixed optimum network matched-filter SNR, $\Ro_s = 6$.
		
		\subsection{\it Performance comparison for fixed injection}
\begin{figure*}
 			\centering
 			\subfloat{\includegraphics[width=1.\textwidth]{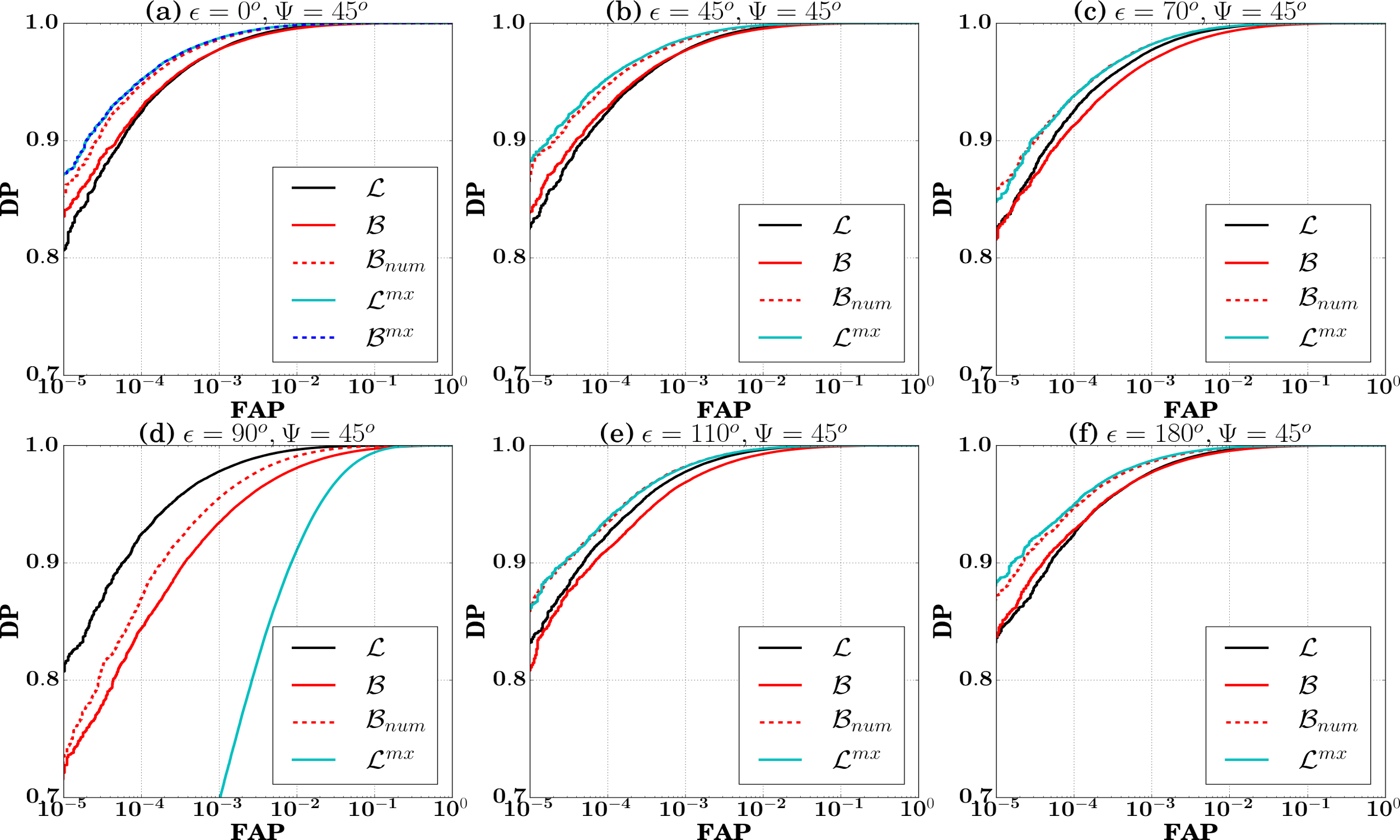}}
 			\caption{\label{Roc1}ROC plots of the  statistics corresponding to a network LHV for fixed injections with different values of $\epsilon$. The signal with SNR, $\Ro_s = 6$ is from $(2 - 10 M_\odot)$ NS-BH system optimally located at $(\theta=140^\circ,\phi=100^\circ)$ with an arbitrary polarization angle $\psi = 45^\circ$. We assume "zero-detuning, high power" Advanced LIGO PSD\cite{aLIGOSensitivity} for all detectors. As one can see in panel (a), ROC curve of $\Bc^{mx}$  overlaps with that of $\Lc^{mx}$. Thus we omit ROC curve of $\Bc^{mx}$ from the remaining panels.}\label{ROC1}
 		\end{figure*}

		In Fig.\ref{ROC1}, we have plotted the ROC curves corresponding to MLR based  statistics $\Lc$,  $\Lc^{mx}$ and  Bayesian statistics $\Bc(X|\Pi_{ph})$, $\Bc^{mx}(X|\Pi_{ph})$ (both analytical [solid curve] and numerical [dashed curve]) for fixed signal with SNR $\Ro_s = 6$ optimally located at $(\theta=140^\circ,\phi=100^\circ)$. The simulations are performed for  6 binary inclination angles namely $\epsilon = 0^\circ, 45^\circ, 70^\circ, 90^\circ, 135^\circ, 180^\circ$ and  the ROC curve are shown in panels (a), (b), (c), (d), (e), (f)  of Fig.(\ref{Roc1}) respectively.  For all cases the binary polarization angle, $\Psi$ is fixed to be equal to $45^\circ$. The numerical Bayesian statistics are obtained by numerically integrating likelihood ratio using Monte-Carlo method with $10^6$ random draws.
		
		For drawing ROC curves, we have taken $2 \times 10^6$ Gaussian noise realizations. For each noise realization, all the statistics are computed with and without signal injection.  For each statistic the FAP and DP are computed  for different threshold values by counting number of times each  statistic crosses that threshold value  when the data contains only the noise as well as when the data contains signal plus noise respectively.
		
		For all inclination angles the ROC curves of generic MLR statistic $\Lc$ looks similar. On the other hand, the {\it hybrid}  MLR statistic $\Lc^{mx}$ has a very strong preference  near face-on/off region. In \cite{Haris:2015iin}, we  show that the $\Lc^{mx}$   improves over $\Lc$ in a wide range of inclination angle except a window $(70^\circ - 110^\circ)$. 
		
		As we have discussed earlier in Sec.\ref{validity},  the analytical approximation of generic Bayesian statistic, $\Bc(X|\Pi_{ph})$ is reasonably valid only in the neighborhood of  $\epsilon = 90^\circ$. We have plotted the ROC curves corresponding to both numerically calculated   Bayesian statistic , $\Bc_{num}$ (dashed red curve) and its analytic approximation, $\Bc$ given in Eq.\eqref{LogB1} (solid red curve). For a fixed  network optimum matched filter SNR $\Ro_s = 6$,  the ROC curves corresponding to  analytic approximation deviates from  $\Bc_{num}$ for all  inclination angles. The detection probability  corresponding to the analytical $\Bc$ always falls below that of $\Bc_{num}$. This is because of low SNR. For high SNR, we expect both ROC curves would  match within a window  of inclination angle around edge-on.
		
		For all inclination angles except $\epsilon = 90^\circ$, the Bayesian statistic (numerical) perform better than the MLR statistic. At edge-on, $\Lc$ performs better than $\Bc$. This is expected, as the $\Lc$ statistic is a Bayesian statistic obtained with an unphysical prior, which is more biased towards edge-on case as described in Sec.\ref{MLRasB} (See Eq.\eqref{Bstat0}). However $\Bc$  statistic is derived for a flat prior on the polarization sphere. Compared to the {\it hybrid}  MLR statistic, Bayesian statistic shows improvement only for edge-on signal. 
		
		The performance of  {\it hybrid}  Bayesian statistic  $\Bc^{mx}$ always  matches with that of  $\Lc^{mx}$. This is because, as one can note in Eq.\eqref{Bstat_face3}, the Bayesian statistic tuned for face-on/off  $\Bc^{0,\pi}$  is equal to $\Lc^{0,\pi}$ with a very small logarithmic correction. As $\Bc^{mx}$ is defined as maximum  of $\Bc^{0}$ and $\Bc^{\pi}$, it is expected that $\Bc^{mx}$ shows the same behavior of $\Lc^{mx}$, which is the maximum  of   $\Lc^{0}$ and $\Lc^{\pi}$. Further the ROC curve corresponding to  analytical approximation of  $\Bc^{mx}$ matches very well with that of numerically evaluated  $\Bc^{mx}$ for all cases. Since the ROC curves of $\Lc^{mx}$ and $\Bc^{mx}$ overlaps very well, in the figures we explicitly plot ROC curve corresponding to  $\Bc^{mx}$ only for the $\epsilon = 0$.
		
		In summary,
			\begin{enumerate}[label=(\alph*)]
				\item Near face-on/off cases, the {\it hybrid} statistics performs better than generic statistics, $\Lc$ and $\Bc$. 
				
				\item  Near edge-on case the generic MLR statistic out performs all other statistics because of the inbuilt unphysical prior, which is skewed towards edge-on case.
			\end{enumerate}

 		\begin{figure*}
 			\centering
 			\subfloat{\includegraphics[width=1.\textwidth]{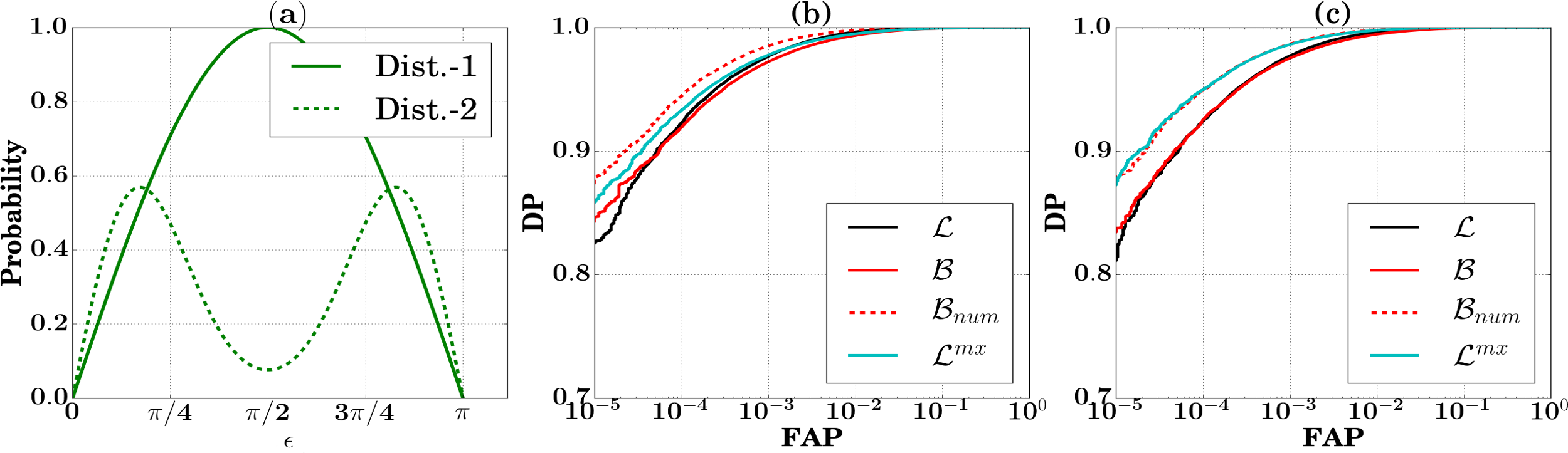}}
 			\caption{\label{Roc2} Panel (a) is the plot of  two sampling distributions of $\epsilon$. Panel (b) gives  ROC plots for 4 different statistics  corresponding to a network LHV when the injected signal's inclination angle, $\epsilon$  drawn from {\bf Dist-1} and panel (c) gives ROC plots for injections with $\epsilon$ drawn from {\bf Dist-2}. In both cases  sky location, and polarization angle are sampled uniformly. The injections are with SNR, $\Ro_s = 6$ and are from $(2 - 10 M_\odot)$ NS-BH system. We assume "zero-detuning, high power" Advanced LIGO PSD\cite{aLIGOSensitivity} for all detectors.}\label{ROC2}
 		\end{figure*}  
			 
		\subsection{\it Performance comparison for injections sampled from a distribution}
		
		In this simulation,the injected binary parameters,  inclination angle $\epsilon$, polarization angle $\Psi$ and source location $(\theta, \phi)$ are sampled from a given distribution. The masses of binary system is fixed to be  $(2 - 10 M_\odot)$ with the  optimum network SNR $\rho_s =6$. The source location $(\theta, \phi)$ is drawn uniformly from celestial sphere and polarization angle $\Psi$ is sampled uniformly from  $[0^\circ, 90^\circ]$. We perform this exercise for two distinct distributions, {\bf Dist-1} and {\bf Dist-2} of inclination angle $\epsilon$.
	
		The {\bf Dist-1}, draws $\cos(\epsilon)$ uniformly from [-1,1] and is denoted by a green (solid) line in panel (a) of Fig.(\ref{Roc2}). As seen, in the figure, the population of random samples drawn from this distribution contains  more of edge-on sources than that of face-on.
		
		In {\bf Dist-2}, the $\epsilon$ follows the distribution proposed in Eq.(28) of \cite{Schutz:2011tw}  [see green (dashed) line in panel (a) of Fig.(\ref{Roc2})].
		\be
		\Pr (\epsilon) = 0.076076 \ll( 1+ 6 \cos^2 \epsilon + \cos^4 \epsilon \rr)^{3/2} \sin \epsilon .
		\ee
		
		The {\bf Dist-2} is a realistic distribution of $\epsilon$, where the   SNR  information is folded in the distribution along with the geometric prior. Since we know that the edge-on sources have less SNR than face-on sources, we expect to see less number of edge-on systems than face-on. As a result, there	would be a dip in the curve (dashed line) with respect to the {\bf Dist -1} (solid line).
		
		Fig.(\ref{ROC2}.b), gives the ROC curves corresponding to   {\bf Dist-1} and panel 	Fig.(\ref{ROC2}.c) gives the ROC curves corresponding to   {\bf Dist-2}.
		
		 For {\bf Dist-1}, the numerical $\Bc$ statistic out performs both generic MLR statistic and {\it hybrid}  statistics. However the ROC curve for analytical approximation of $\Bc$ statistic  is not matching with that of  numerically computed $\Bc$ statistic. For {\bf Dist-2},  ROC curves of $\Bc_{num}$   and {\it hybrid}  statistics overlap well.

		\section{Concussion}\label{conclussion}
		In this article, we address the Bayesian approach for CBC GW search with a multi-detector network with advanced interferometers like LIGO-Virgo.
		
		 We show that the multi-detector MLR statistic obtained by maximizing likelihood ratio over the extrinsic parameters is equal to a Bayesian statistic  with an unphysical prior. Further, we obtain an analytic approximation for alternative  Bayesian statistic with uninformative prior on extrinsic parameters. We also derive Bayesian statistic tuned for face-on/off binaries and construct a {\it hybrid}  Bayesian statistic as a complimentary to {\it hybrid}  MLR statistic devised in \cite{Haris:2015iin}. 
		
		We compare the performance of statistics by means of ROC curves. {\it We observe that  for a wide range of binary inclination angles, the the {\it hybrid}  statistic (both MLR and Bayesian) gives higher detection rates compared to both generic MLR and generic $\Bc$ statistics. Further,  in the neighborhood of $\epsilon = 90^\circ$, where the {\it hybrid}  statistic gives less detection probability, the generic $\Bc$ statistic also fall behind generic MLR statistic  in terms of performance.} 
		
		We performed  the simulations for the network matched filter SNR ($\Ro_s = 6$) due to expressive computation. We notice that  the analytic approximation of Bayesian statistic is not matching  that of numerical Bayesian statistic. This could be due to low SNR value for which we have done the simulation.  As we expect much higher network SNR in real situations, we expect the analytical $\Bc$ is still useful for real searches.

		\section{Acknowledgment}
		The work was supported  by A. Pai’s MPG-DST Max-Planck India Partner Group Grant. The authors availed the 128 cores computing facility established by the MPG-DST Max Planck Partner Group  at IISER TVM. This document has been assigned LIGO laboratory document number LIGO-P1600115.

		\appendix
		
		\section{Relation between  $\{\A,\phi_a,\epsilon,\Psi\}$ and $\{\Ro_L,\Ro_R,\Phi_L,\Phi_R\}$}
		\label{appendix-extrinsic}
		The new parameters, $\{\Ro_L,\Ro_R,\Phi_L,\Phi_R\}$ are related to the physical parameters, $\{\A,\phi_a,\epsilon,\Psi\}$ as below,
		{\small
			\bea
			\Ro_L e^{i \Phi_L} &=&   \A \|\F_+^{'}\| e^{i \phi_a} \ll[\frac{1+\cos^2 \epsilon }{2} \cos 2 \chi + i \cos \epsilon \sin 2 \chi \rr] , \nn  \\
			\Ro_R e^{i \Phi_R} &=&   \A \|\F_\times^{'}\| e^{i \phi_a} \ll[\frac{1+\cos^2 \epsilon }{2} \sin 2 \chi - i \cos \epsilon \cos 2 \chi \rr] . \nn \\ \label{coordinates}
			\eea}
		The absolute values and the phases of the above equations are $\{\Ro_L,\Ro_R,\Phi_L,\Phi_R\}$ and explicitly given in Eq.(B1) of \cite{Haris:2014fxa}.
		
		The determinant of Jacobian of transformation from $\{\Ro_L,\Ro_R,\Phi_L,\Phi_R\}$ to $\{\A,\phi_a,\epsilon,\Psi\}$  is given by,
		
		\bea
		|J| &=&\ll| \det\ll(\frac{\partial \{\rho_L,\Phi_L,\rho_R,\Phi_R\} }{\partial \{\text{A}_0, \phi_a, \cos \epsilon, \Psi\} } \rr)\rr| \nn \\ 
		&=& \frac{2 \|\F_+'\|^2~ \|\F_\times'\|^2~ \ll|\frac{\Ro_L^2 e^{-2i\Phi_L}}{\|\F'_+\|^2} + \frac{\Ro_R^2 e^{-2i\Phi_R}}{\|\F'_\times\|^ 2}\rr|^{3/2}}{\Ro_L ~\Ro_R} \nn \\
		&=&  \frac{\A_0^3 ~\|\F_+'\|^2~ \|\F_\times'\|^2~ (1-\cos^2 \epsilon)^3}{4~ \Ro_L(A, \phi_a, \cos \epsilon, \Psi) ~\Ro_R(A, \phi_a, \cos \epsilon, \Psi)}~. \label{jacobian}
		\eea
		
		\section{Approximation for the  $\Bc(X|\Pi_{ph})$ integration}\label{app-2}
		
		The Eq.\eqref{Bstat_2} gives the integral as,
		{\small  \begin{widetext}
				\bea
				\mathbf{Exp} [\mc{B}(X|\Pi_{ph}I)] &=& C'' \int_0^{\infty} d\rho_L \int_0^{\infty} d \rho_R \int_0^{2 \pi} d \Phi_L \int_0^{2 \pi} d \Phi_R~  \frac{e^{\Lambda(X)}}{|J|}~ \nn \\
				&=& \frac{C''}{\|\F_+'\|  \|\F_\times'\|} \int_0^{\infty}  d\rho_L \int_0^{\infty} d \rho_R \int_0^{2 \pi} d \Phi_L \int_0^{2 \pi} d \Phi_R~ \rho_L \rho_R~ \ll| \frac{\rho_L^2}{\|\F_+'\|^2} e^{2 i \Phi_L} + \frac{\rho_R^2}{\|\F_\times'\|^2} e^{2 i \Phi_R} \rr|^{-\frac{3}{2}} \nn \\
				&& \text{Exp}\ll[\rho_L\hat{\rho}_L \cos (\Phi_L - \hat{\Phi}_L) - \frac{1}{2} \rho_L^2\rr]~ \text{Exp}\ll[\rho_R \hat{\rho}_R \cos (\Phi_R - \hat{\Phi}_R)  - \frac{1}{2} \rho_R^2\rr] ~. \label{Bstat_5}
				\eea
			\end{widetext}}
			
			The denominator is slowly varying along $\Phi_L$ compared to the exponential term in the numerator. If $\rho_L \hat{\rho}_L$ is not small,  then $\text{Exp}\ll[\rho_L\hat{\rho}_L \cos (\Phi_L - \hat{\Phi}_L)\rr]$ will be significant only in a very small range of $\Phi_L$ around $\hat{\Phi}_L$. Thus we can approximate  $\cos(\Phi_L - \hat{\Phi}_L)$ by $1- \frac{(\Phi_L - \hat{\Phi}_L)^2}{2}$. By this expansion the  numerator becomes a Gaussian function in $\Phi_L$ centered at $\hat{\Phi}_L$. Since the denominator is slowly varying function of $\Phi_L$ compared to the numerator, we can  replace $\Phi_L$s in the denominator by $\hat{\Phi}_L$ and then approximate the integral by Gaussian integral. Applying similar argument for $\Phi_R$ integration , the $\Bc$ statistic finally becomes,
			
			{ \small \bea
				\mathbf{Exp} [\mc{B}(X|\Pi_{ph})] = &&\frac{\pi }{C
					\|\F_+'\|  \|\F_\times'\|} \int_0^{\infty}  d\rho_L  \int_0^{\infty}  d\rho_R \sqrt{\frac{\rho_L~ \rho_R}{\hat{\rho}_L \hat{\rho}_R}}  \nn \\ &&  \frac{\text{Exp}\ll[\rho_L \hat{\rho}_L  - \frac{\rho_L^2}{2}\rr] \text{Exp}\ll[\rho_R \hat{\rho}_R  - \frac{\rho_R^2}{2}\rr] }{\ll| \frac{\rho_L^2}{\|\F_+'\|^2} e^{2 i \hat{\Phi}_L} + \frac{\rho_R^2}{\|\F_\times'\|^2} e^{2 i \hat{\Phi}_R} \rr|^{3/2}}~. \nn \\ \label{Integral4}
				\eea}
			
			Please note this approximation is valid only if the denominator of\eqref{Integral4} is non zero.
			
			As one can see in Eq.\eqref{Integral4}, the numerator of the integrand can be converted to a product of Gaussian functions in $\Ro_L$ and $\Ro_R$ centered at $\hat \Ro_L$ and $\hat \Ro_R$ respectively. If $\hat \Ro_L$ and $\hat \Ro_R$ are large enough, by assuming stationarity for the remaining part of the integrand around $(\hat \Ro_L,\hat \Ro_R)$ we can approximate the integrals as,
			
			{\small \bea
				\mathbf{Exp} [\mc{B}(X|\Pi_{ph})]
				= \frac{2 \pi^2  ~e^{\frac{\hat{\rho}^2_L + \hat{\rho}^2_R}{2}}~ \|\F_+'\|^2  \|\F_\times'\|^2}{C \ll| \ll( \|\F_\times'\| \hat{\rho}_L \rr)^2  e^{2 i \hat{\Phi}_L} + \ll( \|\F_+'\| \hat{\rho}_R \rr)^2  e^{2 i \hat{\Phi}_R} \rr|^{3/2}}~. \nn \\  \label{Integral7}
				\eea}
			
			Now the non vanishing condition on the denominator of the integrand reduces to,
			{\small \be
				\ll| \frac{\rho_L^2}{\|\F_+'\|^2} e^{2 i \hat{\Phi}_L} + \frac{\rho_R^2}{\|\F_\times'\|^2} e^{2 i \hat{\Phi}_R} \rr| \neq 0~.
				\ee} 
			In other words, both  $\frac{\rho_L}{\|\F_+'\|} - \frac{\rho_R}{\|\F_\times'\|}$, and  $\hat{\Phi}_L - \hat{\Phi}_R \pm \frac{\pi}{2}$ are not equal to zero simultaneously. This happens when the signal is either face-on or face-off.  
			
			\bibliography{paper}
		\end{document}